\documentstyle[aps,preprint,tighten,floats]{revtex}

\newcommand{\lsim}{\mathrel{\mathop{\kern 0pt \rlap
  {\raise.2ex\hbox{$<$}}}
  \lower.9ex\hbox{\kern-.190em $\sim$}}}
\newcommand{\gsim}{\mathrel{\mathop{\kern 0pt \rlap
  {\raise.2ex\hbox{$>$}}}
  \lower.9ex\hbox{\kern-.190em $\sim$}}}

\begin{document}
\preprint{
\begin{tabular}{r}
DFTT 69/96
\\
JHU--TIPAC--96025
\\
SFB--375/123
\\
November 1996
\end{tabular}
}

\title{Perspectives for Detection of a Higgsino--like Relic Neutralino}

\author{\bf
A. Bottino$^{\mbox{a}}$
\footnote{
\noindent
E--mail: bottino@to.infn.it, fornengo@jhup.pha.jhu.edu,
mignola@vxcern.cern.ch, 
\\
Marek.Olechowski@Physik.TU-Muenchen.DE,
scopel@ge.infn.it
},
N. Fornengo$^{\mbox{b,a}}$,
G. Mignola$^{\mbox{c,a}}$
\footnote{
Present address: LAPP, Annecy, France
},
M. Olechowski$^{\mbox{a}}$
\footnote{On leave of absence from the Institute of Theoretical Physics, Warsaw
University, Poland.
\\
 Present address: Physics Department, Technische Universit\"at M\"unchen,
D-85748 Garching, Germany}
\\
and S. Scopel$^{\mbox{d}}$}
\vspace{1 mm}
\address{
\begin{tabular}{c}
$^{\mbox{a}}$
Dipartimento di Fisica Teorica, Universit\`a di Torino and \\
INFN, Sezione di Torino, Via P. Giuria 1, 10125 Turin, Italy
\\
$^{\mbox{b}}$ Department of Physics and Astronomy,
The Johns Hopkins University,
\\
Baltimore, Maryland 21218, USA.
\\
$^{\mbox{c}}$
Theoretical Physics Division, CERN, CH--1211 Geneva 23, Switzerland
\\
$^{\mbox{d}}$
Dipartimento di Fisica, Universit\`a di Genova and\\
INFN, Sezione di Genova, Via Dodecaneso 33, 16146 Genoa, Italy
\end{tabular}
}
\maketitle

\begin{abstract}
It has been conjectured by Ambrosanio, Kane, Kribs, Martin and  Mrenna  
(AKM) that 
 the CDF event
$p \bar p \rightarrow e^+ e^- \gamma \gamma \rlap{/} {\rm E_T}$ 
is due to a decay chain involving 
two  neutralino states (the lightest and the next-to-lightest ones). 
The lightest neutralino ($\chi_{AKM}$) has been further considered 
by Kane and Wells as a candidate for cold dark matter. In this paper we 
examine the properties of relic $\chi_{AKM}$'s in their full 
parameter space, and examine the perspectives for detection 
by comparing theoretical predictions to  sensitivities 
of various experimental searches. We find that for most 
regions of the
parameter space the detectability of a relic $\chi_{AKM}$ would require  
 quite substantial improvements in current 
experimental sensitivities. 
The measurements of neutrino fluxes from the 
center of the Earth and of an excess of $\bar p/p$ in cosmic rays are 
shown to offer some favorable perspectives for investigating a region of 
the $\chi_{AKM}$ parameter space around the maximal $\tan \beta$ value allowed
by the model. 
\end{abstract}  
\noindent
\newpage

\section{Introduction}

   The occurrence at CDF of the single event 
$p \bar p \rightarrow e^+ e^- \gamma \gamma \rlap{/} {\rm E_T}$ \cite{cdf} 
has prompted two different supersymmetric interpretations 
\cite{dimo,ambro1,dtw,ln,hty,ambro2}, although also a non-supersymmetric explanation 
has been proposed \cite{moha}. 
   
The two supersymmetric interpretations have a common scheme to 
explain the CDF event: first, an $\tilde e^+ \tilde e^-$ pair is 
produced, $p \bar{p} \rightarrow  
\tilde{e}^+ \tilde{e}^-$, then the following decay chain takes place 
$\tilde e^{\pm} \rightarrow e^{\pm} \tilde X_2, \tilde X_2 
\rightarrow \tilde X_1 \gamma$, where $\tilde X_1$ and $\tilde X_2$ 
are the lightest and the next-to-lightest supersymmetric  
particles (LSP and NLSP, respectively). The two supersymmetric 
interpretations differ in the identification of $\tilde X_1$ and 
$\tilde X_2$ with definite supersymmetric particles. In one interpretation 
\cite{ambro1,ambro2} $\tilde X_1$ and $\tilde X_2$ are identified with the 
lightest and next-to-lightest neutralinos  $\chi_1$ and $\chi_2$, 
respectively;  in the second interpretation 
\cite{dimo,ambro1,dtw,ln,hty} 
$\tilde X_2$ is the lightest neutralino $\chi_1$, whereas $\tilde X_1$
is identified with the gravitino, which has the role of the lightest 
supersymmetric particle. 

To test whether or not one of the supersymmetric interpretations is 
correct, various possible processes which can occur at CERN LEP 
or at Fermilab Tevatron have been discussed in Refs. 
\cite{dimo,ambro1,dtw,ln,hty,ambro2}. 

Furthermore, should one of the supersymmetric interpretations be valid, 
this would have implications for the presence of supersymmetric particles as 
relics in the Universe; these could also provide a substantial contribution 
 to the cosmological matter density \cite{kane}.  Thus the natural question 
arises, whether any of these fossil particles could be detected either 
directly or indirectly (or whether they are already excluded in force of the 
present experimental bounds).

The case of relic gravitinos would practically represent a hopeless situation
for experimental investigation, since gravitino interactions 
are too weak to allow detection \cite{ellis}. In models of gauge-mediated 
supersymmetry breaking (where the gravitino is the LSP) 
also the lightest messenger particle may be a viable dark matter candidate with 
a substantial relic abundance. This possibility has been recently
investigated in Ref.\cite{dgp}. We do not pursue the discussion of the 
gravitino case any further here.

Instead, in this paper we address the problem of the possible 
detection of relic neutralinos, whose specific properties are appropriate 
to a correct interpretation of the CDF event \cite{kane}.

As discussed in Refs.\cite{ambro1,ambro2}, the interpretation of the CDF 
event in 
terms of the decay chain $p \bar p \rightarrow \tilde e^+ \tilde e^-$, 
$\tilde e^{\pm} \rightarrow e^{\pm}  {\chi}_2,  {\chi}_2 
\rightarrow  {\chi}_1 \gamma$, sets a number of very stringent 
constraints 
on the supersymmetric parameter space, and in particular on the 
nature of the neutralinos. These constraints, due to the kinematics of the 
event, 
and to the required sizes for the relevant cross section and decay 
branching ratios, 
imply that ${\chi}_1$ and ${\chi}_2$ are a very pure higgsino and a 
very pure photino, respectively. 
Detailed descriptions of the resulting supersymmetric parameter space 
are given in Ref.\cite{ambro2}, and some of these results will also be 
reported here in the next section. For the moment, let us just anticipate 
 some of the most prominent features of the model. 
The parameter $\tan \beta$ is in the 
very low side  (i.e. $1 \lsim \tan \beta \lsim 3$) of its 
natural range: $1 \lsim \tan \beta \lsim 50$; also, for the soft-breaking 
gaugino masses one has $M_1 \simeq M_2$, rather than the usual relationship 
$M_1 \simeq M_2/2$, motivated by unification assumption at $M_{GUT}$ 
(definitions and conventions for the supersymmetric parameters are as in 
Ref. \cite{haber}). 
Furthermore, 
the kinematics of the CDF event (combined with the lower bound from LEP 
data)
entails that $m_{{\chi}_2} - m_{{\chi}_1} \gsim 
30$ GeV and 30 GeV $\lsim m_{{\chi}_1} \lsim 65$ GeV. 
In the following, a neutralino eigenstate of the lightest mass, $\chi_1$, 
will be 
denoted 
by $\chi_{AKM}$, when its properties are those required by the 
supersymmetric 
interpretation of the CDF event, as suggested and described 
in Refs.\cite{ambro1,ambro2} (LSP neutralino scenario).

In ref.\cite{kane} some properties of a $\chi_{AKM}$ neutralino as a 
candidate for Cold 
Dark Matter (CDM) were discussed, and the perspectives for a direct detection 
were 
analysed, under the hypothesis that the contribution of $\chi_{AKM}$ to 
the 
cosmological 
density $\Omega$ is substantial.  The analysis was pursued there in the 
extreme case of a 
pure higgsino composition and in general for a parameter space sizably 
narrower than the one allowed by the supersymmetric interpretation of the CDF 
event. 
In particular, it was concluded that the perspectives 
for a 
direct detection for this candidate are rather favorable. 

In the present paper we reconsider the properties of  $\chi_{AKM}$ as 
a relic particle, by 
expanding the previous 
analysis in many instances \cite{bott}:
i) we explicitly take into account a possible 
gaugino-higgsino mixing in $\chi_{AKM}$, which, although very tiny, might   
nevertheless have sizeable consequences in some 
processes for $\tan \beta \simeq 3$; 
 ii) we relax the requirement that $\chi_{AKM}$ contributes 
to $\Omega$ significantly, since we wish to fully explore the 
experimental chances to detect a relic $\chi_{AKM}$, even in the case 
it is not the main component of Cold Dark Matter (CDM);
iii) we implement all the constraints from 
accelerators, 
including $b \rightarrow s + \gamma$
 and the new bounds implied by the LEP2 measurements \cite{LEP2};  
iv) we examine what are the chances to detect relic 
$\chi_{AKM}$'s using various 
detection strategies (direct detection as well as indirect measurements: 
neutrinos from the Sun \cite{fk}  and from the Earth and 
the antiproton/proton ratio 
in cosmic rays).

  The motivations for the previous points are the following. 
  On very general grounds one expects that a higgsino-like neutralino, 
such as  $\chi_{AKM}$, provides a large relic abundance 
$\Omega_{\chi} h^2$, but has very little chances to be detectable. 
Indeed, a higgsino interacts with matter through spin-dependent 
effects, whereas the sensitivities of the experimental searches which 
are based on 
neutralino-matter scattering, however expected to substantially  
improve in the near future, will still remain for a while only 
at the level of the much larger coherent effects \cite{foot}. 
Thus the perspectives for detection of a relic $\chi_{AKM}$ in 
the near future
appear to be rather gloomy. Nevertheless, the conjecture of 
Refs.\cite{ambro1,ambro2,kane} is very challenging and  
potentially so much far-reaching, that it deserves a more 
careful analysis 
from the point of view of the actual perspectives of detectability. 

   Therefore we have  undertaken the present analysis, with the aim of 
investigating the various circumstances  which could 
provide some better perspectives for experimental exploration of 
at least some  physical region allowed to $\chi_{AKM}$. 
This is why first, 
by taking into account the gaugino-higgsino mixing, even if 
small, we explore the possibility that coherent effects may help in 
providing direct and indirect 
detection rates with more substantial contributions than the ones due 
to the spin-dependent effects. Of course, this cannot occur for configurations 
with 
tan $\beta \simeq 1$, but could happen for neutralino compositions at the 
upper side of the allowed 
range for tan $\beta$, i.e. $\tan \beta \simeq 3$. 
Secondly, we have relaxed the 
constraint that $\chi_{AKM}$ provides substantial relic abundance, which 
in itself sounds rather arbitrary and at the same time forces $\chi_{AKM}$
to stay in a region of the parameter space where neutralino cross sections 
are automatically small. After all, should one 
 be able to detect 
relic neutralinos compatible with the unique CDF event, this would already 
be a major breakthrough, 
even if these neutralinos do not provide a large 
$\Omega$! 
Finally, apart from the more standard detection 
techniques  for WIMPs (direct detection and detection of neutrinos from 
macroscopic bodies), also the antiproton/proton ratio in cosmic rays 
has been considered. Indeed,  the $\chi_{AKM}$ neutralino 
holds some features which 
could favor this kind of signal: a small mass and 
a very tiny (but not vanishing) 
mixing \cite{antip}.

    A few more comments are in order here. All experimental data 
have to be correctly implemented in shaping the allowed parameter space 
for $\chi_{AKM}$. Therefore we have taken into account 
the $b \rightarrow s + \gamma$ process which is a very constraining bound 
to be implemented in any realistic model (it is not clear 
whether or not it was properly taken into account in the previous analyses 
of Refs. \cite{kane,fk}). 
Furthermore we have included in our analysis the very recent 
data from LEP2 \cite{LEP2}.  In the next section we show how these new results 
further constrain the parameter space of Refs.\cite{ambro1,ambro2}. 
The bounds on the Higgs masses, which are important for our evaluation of the
detection signals, are obtained from the experimental data of 
Refs.\cite{al,ac}. 

The scheme of this paper is as follows. In section 2 we define the 
supersymmetric parameter space and examine some general properties of 
$\chi_{AKM}$. In sections 3 and 4 we present our results for the  
$\chi_{AKM}$ relic abundance and for its detection rates, respectively. 
Conclusions are finally reported in section 5.

\section{Model parameter space}

Our parameter space has been modelled according to the one of 
Ref.\cite{ambro2}. 
It issues from the requirement that the CDF event is due to the 
process: 
$p \bar p \rightarrow \tilde e^+ \tilde e^-$, 
$\tilde e^{\pm} \rightarrow e^{\pm}  {\chi}_2,  {\chi}_2 
\rightarrow  {\chi}_1 \gamma$. We only consider the case of 
$\tilde e_L$ production, which appears to be the favourite scheme among those
suggested in Ref.\cite{ambro2}. 

The parameters are: $M_1, M_2, \mu, \tan\beta, m_A$ (mass of the CP-even 
Higgs neutral boson), 
$m_{\tilde l} = m_{\tilde q}$ (this is the common mass for sleptons and 
squarks, taken to be  degenerate, with the exception 
of the left-handed selectron (of mass $m_{\tilde e_L}$) and of the lightest 
stop (of mass $m_{\tilde t_1}$) and $\theta_{\tilde t}$ (mixing angle in the 
stop mass matrix). 

Our analysis of the properties of a relic $\chi_{AKM}$ has been performed by
varying the supersymmetric parameters of a low--energy Minimal 
Supersymmetric extension of the Standard Model (MSSM) in the following 
ranges 
\cite{ambro2}:

region A

\begin{eqnarray}
1.05 &\leq {\rm tan} \beta &\leq 1.5 \nonumber \\
55\;\mbox{GeV} &\leq M_2 &\leq 90\; \mbox{GeV} \nonumber \\
0.8 &\leq M_2/M_1 &\leq 1.2 \nonumber \\
- 70\; \mbox{GeV} &\leq \mu &\leq -33\; \mbox{GeV} \nonumber \\
75\; \mbox{GeV} &\leq m_{\tilde e_L} &\leq 140\; \mbox{GeV} \nonumber \\
m_A &=& 60,\; 100,\; 200,\; 400\; \mbox{GeV} \nonumber \\
m_{\tilde q} &=& 250,\; 500,\; 1000\; \mbox{GeV} \nonumber \\
150\; \mbox{GeV} &\leq m_{\tilde t_1} &\leq m_{\tilde q} \nonumber \\
-\pi/2 &\leq \theta_{\tilde t} &\leq \pi/2
\label{eq:setI}
\end{eqnarray}

region B

\begin{eqnarray}
1.5 &\leq {\rm tan} \beta &\leq 2.8 \nonumber \\
40 \;\mbox{GeV} &\leq M_2 &\leq 130 \;\mbox{GeV} \nonumber \\
1.2 &\leq M_2/M_1 &\leq 2 \nonumber \\
- 70 \;\mbox{GeV} &\leq \mu &\leq -33\; \mbox{GeV} \nonumber \\
75\; \mbox{GeV} &\leq m_{\tilde e_L} &\leq 140\; \mbox{GeV} \nonumber \\
m_A &=& 60, \;100, \;200, \;400 \;\mbox{GeV} \nonumber \\
m_{\tilde q} &=& 250, \;500, \;1000 \;\mbox{GeV} \nonumber \\
150\; \mbox{GeV} &\leq m_{\tilde t_1} &\leq m_{\tilde q} \nonumber \\
-\pi/2 &\leq \theta_{\tilde t} &\leq \pi/2
\label{eq:setII}
\end{eqnarray}

In both cases  $m_{\tilde e_L}$ has been further required to satisfy the
kinematical constraints  among $m_{\tilde e_L}$ and the neutralino mass
eigenvalues \cite{ambro2}.

As usual, any neutralino mass--eigenstate is written as a linear superposition 

\begin{equation}
\chi_i = a_i \tilde \gamma + b_i \tilde Z + c_i \tilde {H_s} 
+d_i \tilde {H_a}
\label{eq:neu}
\end{equation}

\noindent
where $\tilde \gamma, \tilde Z$ are the photino and zino states and 
$\tilde {H_s}, \tilde {H_a}$ are defined by 
$\tilde {H_s} = {\rm sin} \beta \tilde H_1^{\circ} + 
{\rm cos} \beta \tilde H_2^{\circ}$, 
$\tilde {H_a} = {\rm cos} \beta \tilde H_1^{\circ} 
- {\rm sin} \beta \tilde H_2^{\circ}$, in terms of the higgsino fields 
$\tilde H_1^{\circ}$, $\tilde H_2^{\circ}$, supersymmetric partners of the Higgs
fields $H_1^{\circ}$, $H_2^{\circ}$, which provide masses to the down--type 
and up--type quarks, respectively.

The lightest neutralino state $\chi_1$, obtainable by varying the 
supersymmetric parameters in the regions A and B is what we 
define as a $\chi_{AKM}$ neutralino. Its mass turns out to be confined in 
the range: 30 GeV $\lsim m_{\chi} \lsim$ 65 GeV (in region B the upper limit is
about 60 GeV). 

It is worth noticing here that in the previous analyses of relic 
$\chi_{AKM}$'s \cite{kane,fk}, only a restricted  region of the parameter space 
($\tan \beta \simeq 1$)
was considered, where 
higgsino purity in $\chi_{AKM}$ is most pronounced. Also, in Refs. 
\cite{kane,fk} only the lowest 
part of the neutralino mass range was considered, 30 GeV $\leq m_{\chi} \leq$ 
40 GeV, in order to avoid the Z-pole (and possibly Higgs-poles) in the 
neutralino pair-annihilation 
 cross section, where the evaluation of the neutralino 
relic density requires great care. In this paper we include in our 
discussion both region A and region B of supersymmetric parameters. 
We also consider the whole $m_{\chi}$ range and  discuss the effect on the 
detection signals of a careful calculation of the relic abundance over the 
poles of the neutralino-neutralino annihilation cross section. 

On the other side, as previously mentioned, the new constraints from  LEP2 
\cite{LEP2} have been included. We show some effects of these constraints 
in Fig.1. This figure displays in the plane $\mu-M_2$  those AKM 
configurations of regions $A$ and $B$ (see Eqs.(\ref{eq:setI},\ref{eq:setII})) 
which survive the $b \rightarrow s + \gamma$ constraint. It turns out that 
some of them are 
already excluded
by the new LEP data. The various curves denote the chargino isomass contours 
at fixed $\tan \beta$ which correspond to the current LEP lower bound on the
chargino mass \cite{LEP2}.  For a given 
value of $\tan \beta$, the configurations 
on the right of the relevant line are disallowed. In particular, one
sees that no AKM configuration survives for $\tan \beta \gsim 2.6$. 

The composition of  $\chi_{AKM}$ is what establishes the size of the neutralino
relic abundance and of the detection rates. This composition is shown in 
Fig.2. In Sect.a of this figure we give the values of the 
weights 
$|a_1|^2, |b_1|^2, |c_1|^2, |d_1|^2$ of the 
$\tilde \gamma, \tilde Z, \tilde H_s, \tilde H_a$ 
components for the smallest value of $\tan \beta$: $\tan \beta = 1.05$. 
In Sect.b we display the values of the same quantities cumulatively for the two
values 
$\tan \beta = 2.15, 2.5$.  We notice that, as
anticipated,  $\chi_{AKM}$ is largely dominated by $H_s$, with a 
next-to-leading contribution from $\tilde Z$. In Sect.c of Fig.2 we give 
a scatter plot for the fractional weights of these two main components of 
$\chi_{AKM}$ over the full grid of 
Eqs.(\ref{eq:setI}, \ref{eq:setII}). 
For some configurations of region B the value of the ratio 
$|b_1|^2$/$|c_1|^2$, even if small, may nevertheless be sizeable enough to 
allow coherent effects in neutralino-matter interaction to overcome the    
spin-dependent ones.

\section{Relic abundance of $\chi_{AKM}$}

The neutralino relic abundance $\Omega_{\chi} h^2$ is evaluated using the
standard formula 

\begin{equation}
\Omega_{\chi} h^2=3.3 \times 10^{-38} 
\frac {1}{\sqrt{g_{*}(x_f)}} \frac{{\rm cm}^2}{I(x_f)}
\label{eq:omega}
\end{equation}
 
\noindent
where  

\begin{equation}
I(x_f)=\int_0^{x_f} dx  <\sigma_{ann} v> .
\label{eq:i}
\end{equation}

\noindent
$<\sigma_{ann} v>$ is the thermally-averaged annihilation cross section times 
the relative velocity, $g_*(x_f)$ is the number of degrees of freedom at the
freeze-out temperature $T_f$ and $x_f = T_f/m_{\chi}$.

 Whenever the neutralino-neutralino 
annihilation cross section is not in the proximity of a pole or when 
anyway a great accuracy in the estimate of $\Omega_{\chi} h^2$ 
is not important, we simply expand  $<\sigma_{ann} v>$ at small velocities 
$<\sigma_{ann} v> = a + bx $ ($x = T/m_{\chi}$), 
and thus $I(x_f) = ax_f + b{x_f}^2/2$ \cite{ehn,swo}. 

Otherwise, when we are close to a pole for the annihilation cross section
 and we require a careful evaluation for the relic abundance, 
the function $I(x_f)$, which entails multiple integrations over $x$
and over the two particle velocities, is carefully evaluated, in part 
analytically 
and in part numerically \cite{gs,gg}. Since this procedure is much 
 computer-time consuming, we have applied the following selection criteria. 
Out of the full set of neutralino configurations explored through a scanning 
of the regions A and B,  we have selected a number of 
configurations (denoted as set S in the following)  which, according to our 
estimates of detection rates, have more chances to be detected in the 
future. The set S will be precisely defined later on. 
For the configurations of set S the relic abundance has been evaluated 
in the exact way (and compared to the approximated estimate), whereas for other 
configurations only the approximate method, based on the 
 low-velocity expansion, has been adopted.

In the evaluation of $<\sigma_{ann} v>$ all the $f \bar f$ final states as
well as the complete set of Born diagrams have been taken into account 
\cite{bda}.

In Fig.3 we display the results of our evaluation. 
Diamonds and crosses represent the values 
of $\Omega_{\chi} h^2$ for configurations of set S (diamonds denote the values 
of $\Omega_{\chi} h^2$ calculated in the exact way, crosses give the values 
obtained with the low-velocity approximation).  Dots denote the values 
$\Omega_{\chi} h^2$ for the other configurations (in 
the low-velocity approximation).

Some interesting  features show up in this figure: i) $\Omega_{\chi} h^2$ 
displays the typical  dip at about 45 GeV (Z-pole); ii) in going through the 
pole in $\sigma_{ann}$, the approximated value of $\Omega_{\chi} h^2$ 
changes from an overestimate to an underestimate of the correct value (see 
Sect.b of Fig.3); iii) as expected, in region A 
($\tan \beta \simeq 1$) the relic abundance may be quite sizeable, and may 
even fall in the favorite range 
$\Omega_{CDM} h^2 \simeq 0.2 \pm 0.1$ \cite {bbe}, whereas in region B it turns
out to be systematically below 0.03.

The evaluation of $\Omega_\chi h^2$ is important here not only to
establish the role played by the $\chi_{AKM}$ neutralino as a CDM candidate, but
also to provide the value of the local (solar neighborhood) density 
$\rho_{\chi}$. This quantity enters in all the detection rates to be  
considered in the following. 
Here, to determine the value of $\rho_\chi$, we adopt  the following rescaling 
recipe \cite{gaisser}. For each point of the parameter
space, we take into account the relevant value of the cosmological neutralino
relic density. When $\Omega_\chi h^2$ is larger than a minimal
$(\Omega h^2)_{min}$, compatible with observational data and with large-scale 
structure calculations, we simply put $\rho_\chi=\rho_l$.
When $\Omega_\chi h^2$ turns out  to be less than $(\Omega h^2)_{min}$, 
and then the neutralino may only provide a fractional contribution
${\Omega_\chi h^2 / (\Omega h^2)_{min}} \equiv  \xi$
 to $\Omega h^2$, we take $\rho_\chi = \rho_l \xi$.
The value to be assigned to $(\Omega h^2)_{min}$ is
somewhat arbitrary, in the range 
$0.03 \lsim (\Omega h^2)_{min} \lsim 0.2$. In the present paper we have used 
$(\Omega h^2)_{min} = 0.03$. As far as the value of $\rho_l$ is concerned, we
have taken the  representative value $\rho_l = 0.5~{\rm GeV \cdot cm^{-3}}$. 
This corresponds to the
central value of a recent determination of $\rho_l$, based on 
a flattened dark matter distribution and microlensing data: 
$\rho_l = 0.51_{-0.17}^{+0.21}~{\rm GeV \cdot cm^{-3}}$  \cite{turner}.

\section{Detection rates for $\chi_{AKM}$}

The most natural question to be asked now is whether there may be some chance
to detect a relic neutralino with the properties of $\chi_{AKM}$. To provide an
answer to this question we examine in detail three of the main methods for
detecting relic particles (neutralinos in our case) \cite{jkg}: 
i) direct detection, 
ii) detection of neutrinos from macroscopic bodies (Earth and Sun), iii)
measurement of an excess  of $\bar p/p$ in cosmic rays, due to neutralino-neutralino 
annihilation in the halo. 

As mentioned in the introduction, on  very general grounds one expects that 
the interaction of the $\chi_{AKM}$ neutralino with matter takes place 
through spin-dependent effects \cite{kane}. This is due to the fact that 
$\chi_{AKM}$ is an almost pure higgsino, and then couples to quarks
mainly through a $Z$-exchange. This is particularly true for configurations
where $\tan \beta \simeq 1$. However, as shown in Fig. 2, for $\chi_{AKM}$ 
compositions at the upper extreme of the allowed $\tan \beta$ range, 
$i.e. \tan \beta = 2.5$, the higgsino-gaugino mixing parameter 
$|b_1|^2/|c_1|^2$ can reach a level of 
$\simeq  10\%$ and then may switch on some coherent effects through
Higgs-mediated or squark-mediated processes. This effect may trigger
an enhancement in direct detection rates as compared to a simple evaluation
based on spin-dependent effects only. A second beneficial effect due to a 
mixing in $\chi_{AKM}$ is that also the neutrino outcome from the Earth 
(typically increased by coherent effects) might  be sizably enhanced.

A third detection method for $\chi_{AKM}$ investigated in the present paper is
the measurement of the antiproton component in cosmic rays. This experimental
mean is very interesting, in view of the upcoming projects \cite{pam,ams},
which should substantially increase the number of measured antiprotons, bringing
the present total number of about 30 to something of order 600 in a few-year 
time \cite{ams}. This remarkable increase in statistics should soon allow us to 
discriminate between a fast-varying 
spectrum of secondary antiprotons and a flat spectrum of antiprotons of 
exotic origin for kinetic energies in the range
$100\,\,\mbox{MeV} \lsim T \lsim$ a few GeV. Also from the theoretical point of view the peculiarity of 
$\chi_{AKM}$ offers some interesting features for the $\bar p/p$ ratio 
(tiny higgsino-gaugino mixing and small mass).

Now let us define our set S of AKM configurations. This is a set of
representative points within regions A and B which satisfy the following
prerequisites: their predicted signals either for detection of neutrinos from
Earth and Sun (at least one of these) or for detection of antiprotons in cosmic
rays is within two orders of magnitude from the current value of the relevant
experimental upper bound (at 90 \% C.L.).

\subsection{Direct detection}

Let us start our analysis of the detection methods from the most natural one:
direct detection. This consists in the measurement of the energy released by 
a neutralino in its scattering off a nucleus in an appropriate detector, 
by using very different experimental techniques \cite{jkg}. Some of the most
recent experimental results are given in 
Refs.\cite{heidelberg,boulby,rome,milano,zaragoza}.

In general, it is expected that, in the neutralino-nucleus scattering, coherent
effects, when allowed by the neutralino composition,  overcome 
spin-dependent effects. 
When this is the case, then the best way to compare experimental data, which
usually refer to a variety of nuclear compositions, is to convert the upper 
limits on
the energy spectra into upper bounds on the neutralino-nucleon scalar 
cross section $\sigma_{scalar}^{(n)}$. 
This procedure is a model-independent one, i.e. it does not depend on the 
neutralino composition. 

However, as previously discussed,  for configurations of region A, 
the $\chi_{AKM}$ is a 
neutralino  with 
 a  high higgsino-purity and then its spin-dependent interactions with matter 
are important. 
Therefore for configurations of region A 
the previous procedure is not
the most appropriate one and consequently we 
 consider a rate  rather than the neutralino-nucleon cross section. 
As far as configurations of region B are  concerned, it turns out that the 
maximal signals are 
already slightly dominated by coherent effects. Therefore, for our comparison 
between predictions and experimental upper limits for configurations of region
B we use both quantities: rates
and cross sections.

For our evaluation of cross sections and rates for the process at hand, we used
the method described in Refs.\cite{direct}. Our results are 
reported in Figs.4-5. In Fig.4 we display the predicted values 
of the rate $R_{NaI}$ for the scattering of a neutralino off a 
NaI detector, integrated over  the range 3.75 KeV $\leq E_{ee} \leq$ 5.25 KeV, where 
$E_{ee}$ is the electron-equivalent energy. The reason for considering this
quantity is that it provides 
one of the most stringent experimental upper bounds 
(for neutralinos interacting through spin-dependent effects and with a mass in 
the range  40 GeV $\leq m_{\chi} \leq$ 75 GeV): 
$R_{NaI}^{expt} \lsim 1$ event/(Kg day) \cite{boulby}. From Fig.4 it
is clear that all the predicted values for AKM configurations fall far below 
the current experimental bound (by more than two orders of magnitude for
both regions, region B being slightly better than region A). 

Fig.5 displays the scatter plot of the neutralino-nucleon cross-section times
the rescaling factor $\xi$
for configurations of region B only and compares
these to the experimental upper bound. The experimental limit shown in this 
figure refers to an experiment using 
a Ge-detector \cite{zaragoza}. This bound is somewhat more restrictive than the
previous one from the NaI-detector, since now we start dealing with coherent
effects and then we can optimize all experimental data to 
obtain the most stringent upper limit. 
 Consequently, the maximal predicted signal for some configurations 
turns out to be a little closer to the current limit,  but however away by 
about two orders of magnitude. 
We notice that some increase in $R_{NaI}$ and $\xi \sigma^{(n)}_{scalar}$ is
due to the refined evaluation of the neutralino relic abundance (see Figs.
4b-5).

\subsection{Neutrinos from the Earth and from the Sun}

Let us turn now to the possible signals consisting of fluxes of up-going muons 
through a neutrino telescope generated by neutrinos produced by pair
annihilations of neutralinos captured and accumulated inside the Earth and the
Sun. The evaluation of the muon fluxes, which is a rather elaborate multistep 
process, has been performed here according to the procedure described in Ref.
\cite{upm}, to which we refer for details. 

In order to conform to the experimental data which we use as upper limits, we
consider here fluxes of up-going muons integrated over muon energies above 
1 GeV. The flux from the Earth $\Phi_{\mu}^{Earth}$ 
is also integrated over a cone of half aperture of $30^{\circ}$ centered at 
the nadir, the one from the Sun $\Phi_{\mu}^{Sun}$ is integrated over the whole
possible outcome from the Sun, i.e. integrated over $25^{\circ}$ around the Sun
direction. We compare our evaluations to the Baksan upper limits: 
$\Phi_{\mu}^{Earth} \leq 2.1 \times 10^{-14} {\rm cm^{-2} s^{-1}} (90 \% C.L.)$, 
$\Phi_{\mu}^{Sun} \leq 3.5 \times 10^{-14} {\rm cm^{-2} s^{-1}} (90 \% C.L.)$ 
\cite{bbb}.

Our results are shown in Figs.6-7.  We notice that for region A (Fig.6a) 
the maximal 
value of $\Phi_{\mu}^{Sun}$, provided only by very few configurations at 
$m_{\chi} \simeq 65$ GeV, 
is $\simeq 5 \times 10^{-15} cm^{-2} s^{-1}$, anyway below the experimental
upper bound roughly by a factor of 6. These configurations were disregarded in
previous analyses \cite{kane,fk}. Most of the other configurations give 
signals  largely spread over more than three decades. 
When we move from region A to region B (Fig.6b), $\Phi_{\mu}^{Sun}$ increases as
expected, since coherent effects start playing some role in enhancing the
neutralino capture rate by the Sun. Here a significant number of 
configurations have a 
predicted level of $\Phi_{\mu}^{Sun}$ within an order of magnitude from the 
current experimental bound. For region B an even more favorable comparison 
between predictions and experimental sensitivity occurs for 
$\Phi_{\mu}^{Earth}$ (Fig.7). Indeed, the maximal predicted value is  away from the
present experimental limit only by a factor 2. However, it is again apparent
from Fig.7 that the predicted values for $\Phi_{\mu}^{Earth}$ are spread over 
a very wide range of a few decades. 
As expected, the configurations with the highest values for
the flux are those with a light Higgs boson A ($m_A \simeq 100$ GeV) and some 
higgsino--zino mixing. The gap 
in between the two groups of configurations in Fig.7 is indeed due to the
step in $m_A$ used in our sampling of the parameter space.

We again notice that some increase in the level of the fluxes in Fig.6b and 
Fig.7 
is due to the refined evaluation of the neutralino relic abundance. Furthermore
it is worth noticing that some improvements in the comparison of the predicted 
values for $\Phi_{\mu}^{Earth}$ and the experimental data may be obtained
through a more refined analysis of the fluxes in terms of their angular 
distribution \cite{monta}.

\subsection{${\bf \bar p}$/p in cosmic rays}

The annihilation of neutralinos in our halo may generate some 
amount of antiprotons in our Galaxy. A way of discriminating them, against 
the background due to the secondary antiprotons  produced by primary cosmic 
particles with the interstellar medium, is to look at the 
$\bar p/p$ spectrum as a function of the kinetic energy $T$. At small 
  $T$ the $\bar p/p$  ratio due to secondaries increases quickly, as $T$ 
increases, whereas the signal has a flat behaviour. 

Measurements of anti-protons have been going on for quite a long time with 
some conflicting results \cite{jkg}. More recent data \cite{pic,mit} seem to
follow the behaviour of secondaries as evaluated in Ref.\cite{sch}. 
However, a much higher statistics is required to find out whether or not there 
might be a signal of some exotic origin for antiprotons. This very intriguing 
problem should be settled in a few-year time, due to upcoming experiments 
\cite{pam,ams} which are expected to collect a total of about 600 antiprotons
\cite{ams}. 

We have evaluated the $\bar p/p$ ratio in the following way: i) the antiproton 
spectrum, as due to the neutralino pair annihilation in the halo, has been
calculated as in Ref.\cite{antip}, ii) the proton spectrum has been taken from
Ref.\cite{ryan}, iii) the propagation of the two fluxes has been evaluated 
using a leaky box model with an energy-dependent confinement time taken from 
Ref.\cite{lapp}, 
iv) the two spectra have been modulated by employing the procedure of 
Ref.\cite{perko} with the modulation parameter of Ref.\cite{mit} and then 
integrated over
the range $ 250 \;\mbox{MeV} \leq T \leq 1000 \;\mbox{MeV}$ to conform to 
the experimental characteristics of one of the most significant  experimental 
data: 
$\bar p/p = 3.14_{-1.9}^{+3.4} \times 10^{-5}$ 
for $ 250 \;\mbox{MeV} \leq T \leq 1000 \;\mbox{MeV}$ \cite{mit}. To make the 
comparison of our predicted
values with the experimental data more meaningful, we use in the following 
the value $\bar p/p \leq 7.5 \times 10^{-5}$ as 
indicative of a 90 \% C.L. limit for antiprotons of exotic origin. 

We show our results in Fig.8. We notice that for a limited number 
of configurations in
region B (Fig.8b) the predicted signal is rather close to the 
experimental value, 
but for many others the signals are away by orders of magnitude. The
maximal predicted value for $\bar p/p$ is below the upper limit by a factor
3-4. 
Again it turns out that the improvement in the calculation of 
$\Omega_{\chi} h^2$ enhances the expected signal.  In Fig.9 we give a scatter plot of 
$\bar p/p$ versus 
$\Phi_{\mu}^{Earth}$ to show how the same set of 
$\chi_{AKM}$ configurations provide the maximal predicted values for both 
of these two quantities. Fig.10 shows where these configurations are located in
the $\mu-M_2$ plane.

\section{Conclusions}

The conjecture that the CDF event
$p \bar p \rightarrow e^+ e^- \gamma \gamma \rlap{/} {\rm E_T}$ 
is due to a decay chain involving 
two  neutralino states (the lightest and the next-to-lightest ones)
\cite{ambro1,ambro2} is certainly very intriguing, although  great caution is 
in order, because of the existence of a single event of this sort and of the
non-uniqueness in its interpretation \cite{dimo,ambro1,dtw,ln,hty,ambro2,moha}. 
Nevertheless, the interpretation of the event 
suggested in  Refs. \cite{ambro1,ambro2}, if correct, would have so much impact on particle
physics, that 
any possible experimental verification of it should be carefully investigated. 
Obviously, accelerators are the most suitable means for this purpose. 

Also the implications for relic supersymmetric neutralinos deserve 
much attention and experimental investigation. With this target in mind, we
have extended the  analyses of Refs.\cite{kane,fk} in many ways. We have
explored a much wider region in the supersymmetric parameter space than
previously done and we have 
examined a variety of different detection means. 
In such a challenging enterprise of
searching for a relic particle the only winning strategy is the one of
combining as many independent searches as possible.

Let us now summarize some of our results. We have considered three 
detection methods for relic neutralinos: i) direct detection, 
ii) detection of neutrinos from macroscopic bodies (Earth and Sun), iii)
measurement of an excess of $\bar p/p$ in cosmic rays, due to 
neutralino-neutralino annihilation in the halo. 
For all of these search techniques we have evaluated the relevant signals 
exploring the widest allowed parameter space of the $\chi_{AKM}$ neutralino 
and we have compared our results to the current experimental bounds. 
In no case present experimental data are such to provide information on 
a relic $\chi_{AKM}$. All experimental methods require a substantial
improvement in sensitivities before they may be capable of exploring some 
sizeable region of the $\chi_{AKM}$ parameter space. In general, the most 
easily accessible region is the one corresponding to the values of 
$\tan \beta$ close to the upper part of the AKM range 
1.05$\leq \tan \beta \leq 3$ and to the smallest values of $m_A$. 
Unfortunately, as discussed in Sect.2, the new LEP2 data already exclude 
the $\chi_{AKM}$ configurations with $\tan \beta \gsim 2.6$. 

For direct detection the most easily accessible part of the $\chi_{AKM}$ 
parameter space requires a very significant experimental improvement in 
sensitivities of 2-3 orders of magnitude. We emphasize that our conclusion 
is based directly on consideration of neutralino-nucleon cross section, 
and then automatically takes into account the most stringent experimental 
measurements. 

The measurement of fluxes of upgoing muons from the center of the Earth 
appears to be in a much better situation, since some configurations are away 
from the present upper bound by a factor of two. However, an improvement 
of at least one order of magnitude in sensitivity would be necessary for an 
exploration of a significant number of configurations. The most suitable 
detector for this job appears to be MACRO with a muon energy threshold of 
about 1 GeV, whereas large-area neutrino telescopes such as AMANDA and 
NESTOR would not have much chances because of a much higher energy threshold. 

We have shown that some $\chi_{AKM}$ configurations may provide 
an excess of $\bar p/p$ in cosmic rays at a level  that is away from the 
present measured value by a factor 3-4. This fact deserves much attention 
in view of the expected increase of statistics in the upcoming experiments 
in space. 

However, a word of caution is in order here. All our evaluations of signals for
configurations  in region B are very sensitive to the value assigned to the
parameter $(\Omega_{\chi} h^2)_{min}$, which enters in our rescaling of
$\rho_{\chi}$. Here we have used $(\Omega_{\chi} h^2)_{min} = 0.03$, which
roughly corresponds to a minimal value for $(\Omega_{\chi} h^2)_{min}$. If a
larger value of 
$(\Omega_{\chi} h^2)_{min} = f \times 0.03$ ($f$ greater than one) is 
employed, then in region $B$ the estimates for $R$ and $\Phi_{\mu}$ have to be
reduced by a factor $1/f$ and the values of $\bar p/p$ by a factor $(1/f)^2$.
The same considerations apply to those configurations of region $A$, whose
relic abundance is smaller than $(\Omega_{\chi} h^2)_{min}$.

Finally, we point out that in the present paper our aim was to establish a comparison
between the level of the maximal signals due to a $\chi_{AKM}$ neutralino 
and the 
experimental sensitivities currently available or expected in a near future. 
It is obvious that, in view of an actual experimental measurement,  specific
signatures and signal to background discriminations should be carefully 
investigated. This is beyond the scope of the present analysis. 

{\bf Acknowledgements}

We wish to thank Lars Bergstr\"om for interesting discussions. N. Fornengo
gratefully acknowledges a post-doc fellowship of the Istituto Nazionale di
Fisica Nucleare 
and G. Mignola one of  the Universit\`a di Torino. M. Olechowski wishes 
to thank the support of the Polish Committee of Scientific Research and grant
SFB-375-95 (research in astroparticle physics) of the Deutschen
Forschungsgemeinschaft. Partial
support was provided by the Theoretical Astroparticle Network under contract 
No. CHRX-CT93-0120. This work was also 
supported in part by the Research Funds of the Ministero dell'Universit\`a 
e della Ricerca Scientifica e Tecnologica.

\vfill
\eject

{\bf Figure Captions}
\vspace{15 mm}

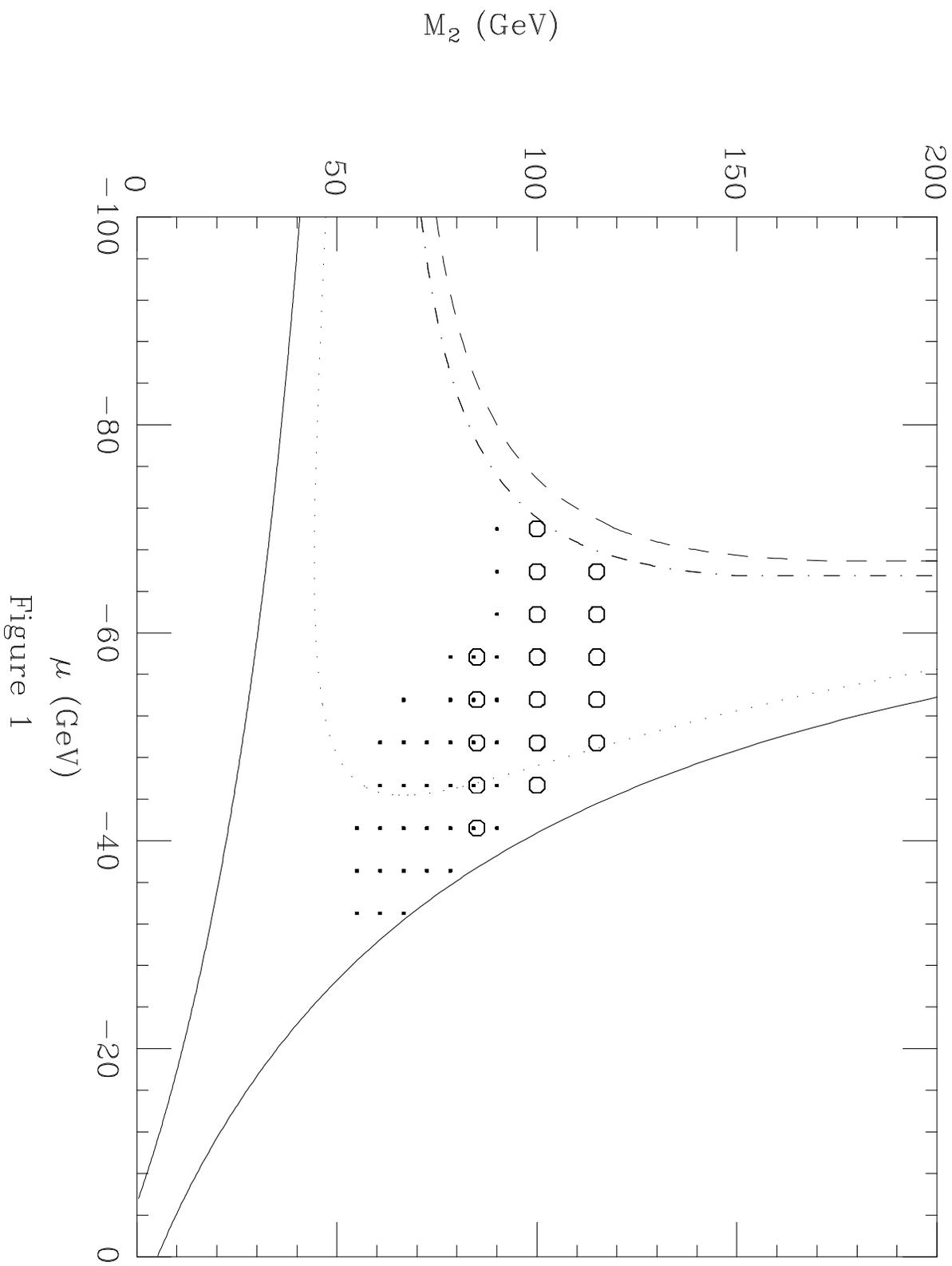
{\bf Figure 1} -- 
AKM configurations of regions $A$ and $B$ 
(see Eqs.(\ref{eq:setI},\ref{eq:setII})) 
which survive the $b \rightarrow s + \gamma$ constraint, displayed in
the plane $\mu-M_2$. The various curves denote the chargino isomass contours 
which correspond to the current LEP lower bound on the chargino mass 
\cite{LEP2}, for different values of $\tan\beta$: 
$\tan\beta = 1.05$ (solid), $\tan\beta = 1.5$ (dotted),
$\tan\beta = 2.6$ (dot--dashed), $\tan\beta = 2.8$ (dashed).
Dots (circles) denote configurations of region A (B).

\vspace{10 mm}

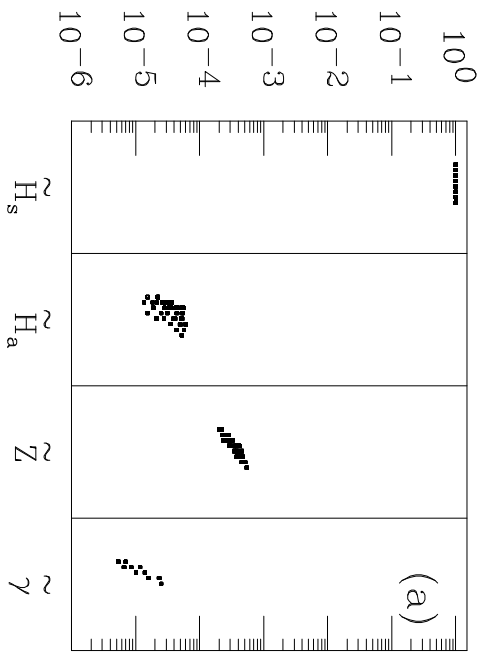
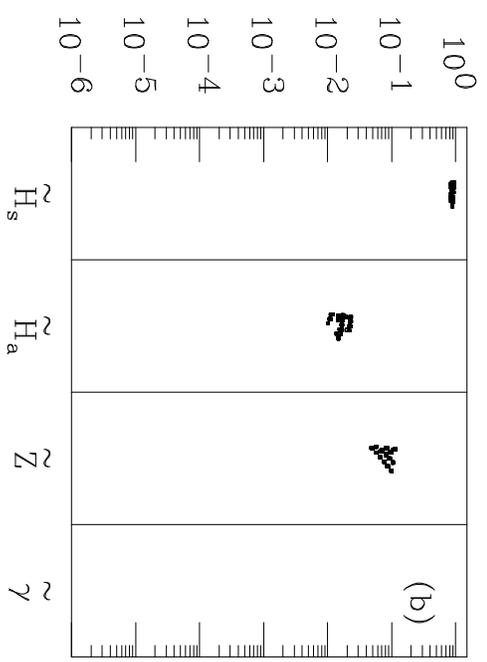
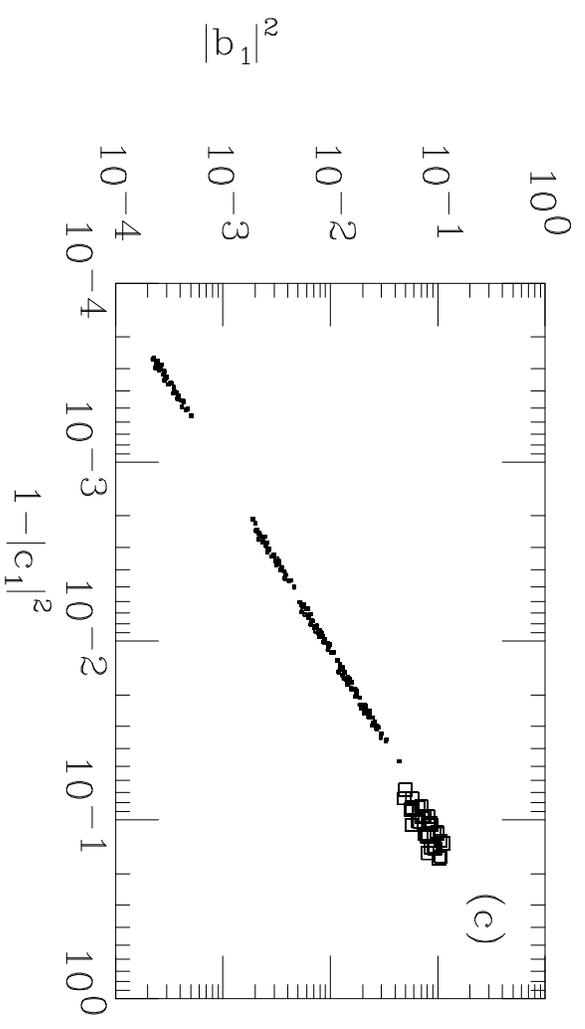
{\bf Figure 2} --
Sect.a displays the values of the weights 
$|a_1|^2, |b_1|^2, |c_1|^2, |d_1|^2$ of the 
$\tilde \gamma, \tilde Z, \tilde H_s, \tilde H_a$ 
components for the smallest value of $\tan \beta$: $\tan \beta = 1.05$. 
Sect.b shows the values of the same quantities cumulatively for the 
values $\tan \beta = 2.15, 2.5$.  Sect.c  shows 
a scatter plot for the fractional weights of the two main components of 
$\chi_{AKM}$ over the full grid of 
Eqs.(\ref{eq:setI}, \ref{eq:setII}).

\vspace{10 mm}

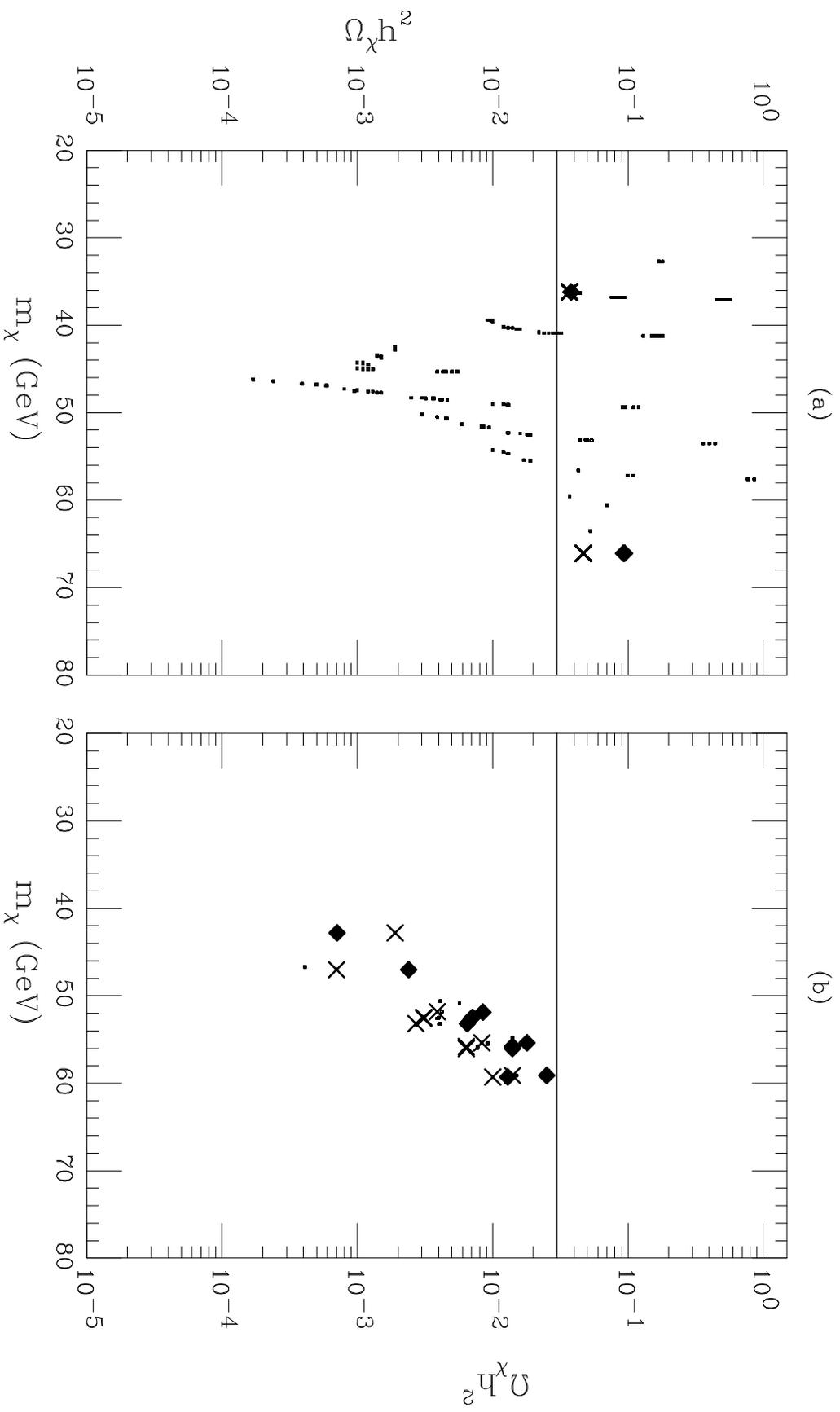
{\bf Figure 3} --
Values of $\Omega_{\chi} h^2$ for configurations in region A (B) are 
given in Sect.a (b).
Diamonds and crosses represent the values 
of $\Omega_{\chi} h^2$ for configurations of set S (diamonds denote the values 
of $\Omega_{\chi} h^2$ calculated in the exact way, crosses give the values 
obtained with the low-velocity approximation).  Dots denote the values 
$\Omega_{\chi} h^2$ for the other configurations (in the low-velocity
approximation only). 
The horizontal line corresponds to the value $\Omega_{\chi} h^2=0.03$, below which 
we apply rescaling for the local neutralino density.

\vspace{10 mm}

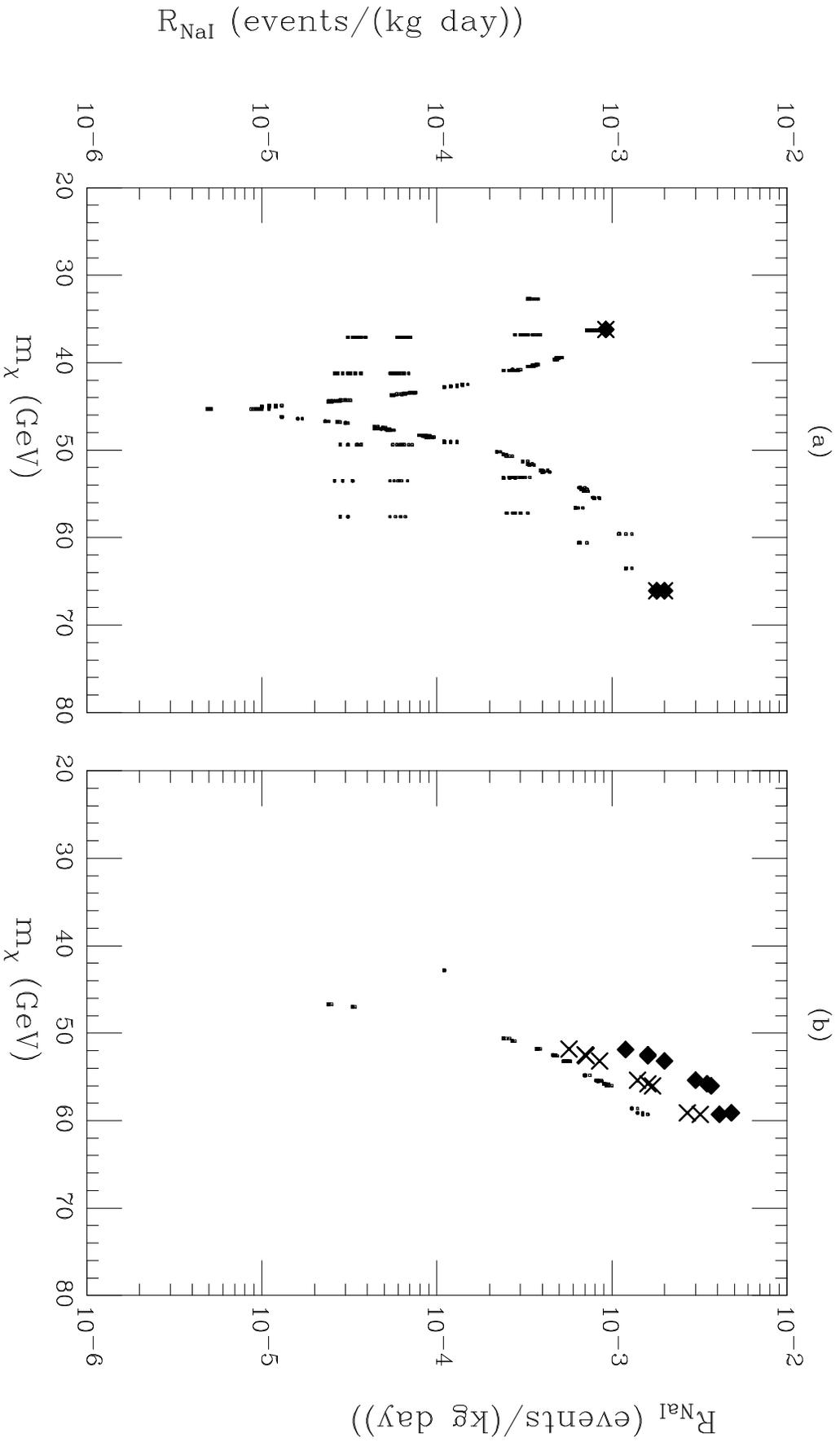
{\bf Figure 4} --
Predicted values 
of the rate $R_{NaI}$ for the scattering of a neutralino off a 
NaI detector, integrated over  the range 3.75 KeV $\leq E_{ee} 
\leq$ 5.25 KeV, where 
$E_{ee}$ is the electron-equivalent energy.
Configurations in region A (B) are given in Sect.a (b).
Diamonds and crosses denote the values of the displayed signal for 
configurations of set S when $\Omega_\chi h^2$ is calculated with the exact
expression and with the low-velocity approximation, respectively. 
Dots denote the values of the displayed signal for the other
configurations, calculated in the low-velocity approximation only.

\vspace{10 mm}

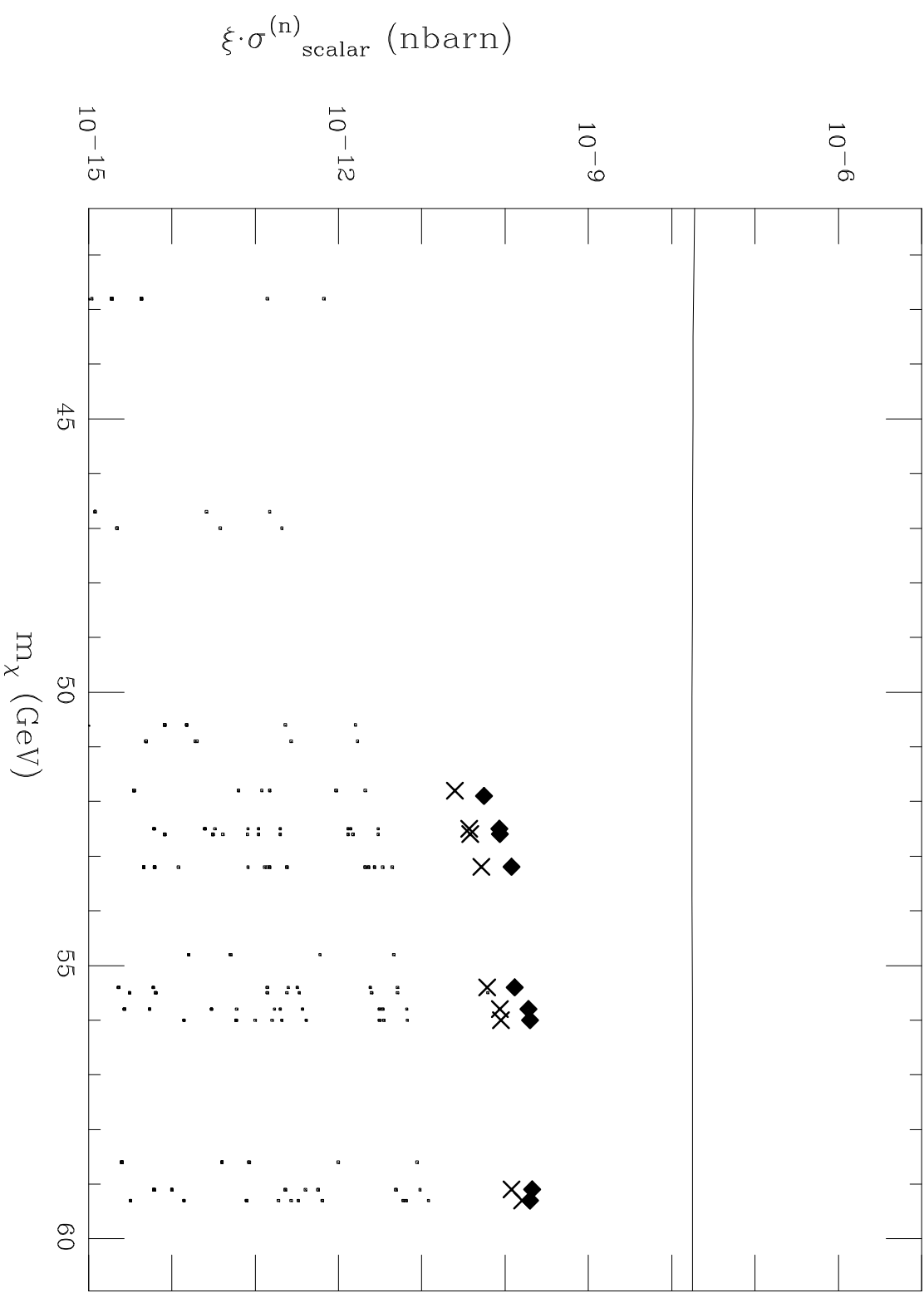
{\bf Figure 5} -- Scatter plot of 
the neutralino-nucleon scalar cross-section times the rescaling factor $\xi$ for
configurations of region B. The line denotes 
the upper bound on the cross-section obtained using 
the experimental results of Ref. \cite{zaragoza}  
(Ge-detector). Diamonds, crosses and dots are as in Fig. 4.

\vspace{10 mm}

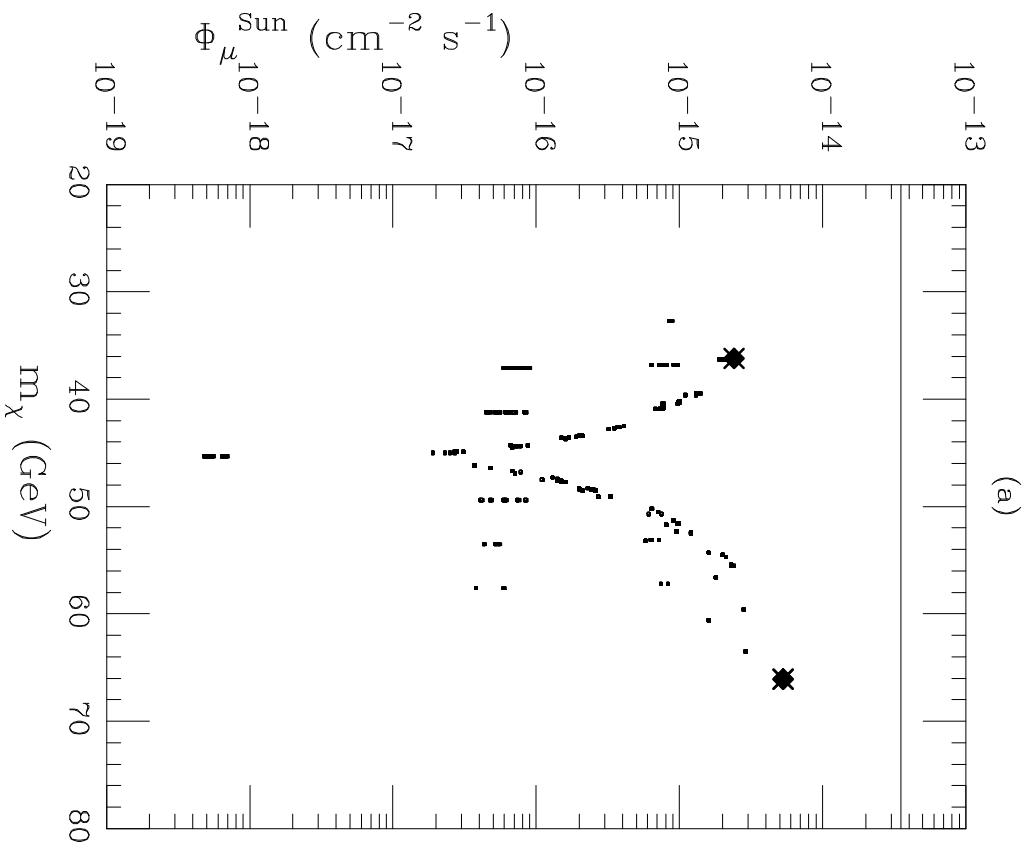
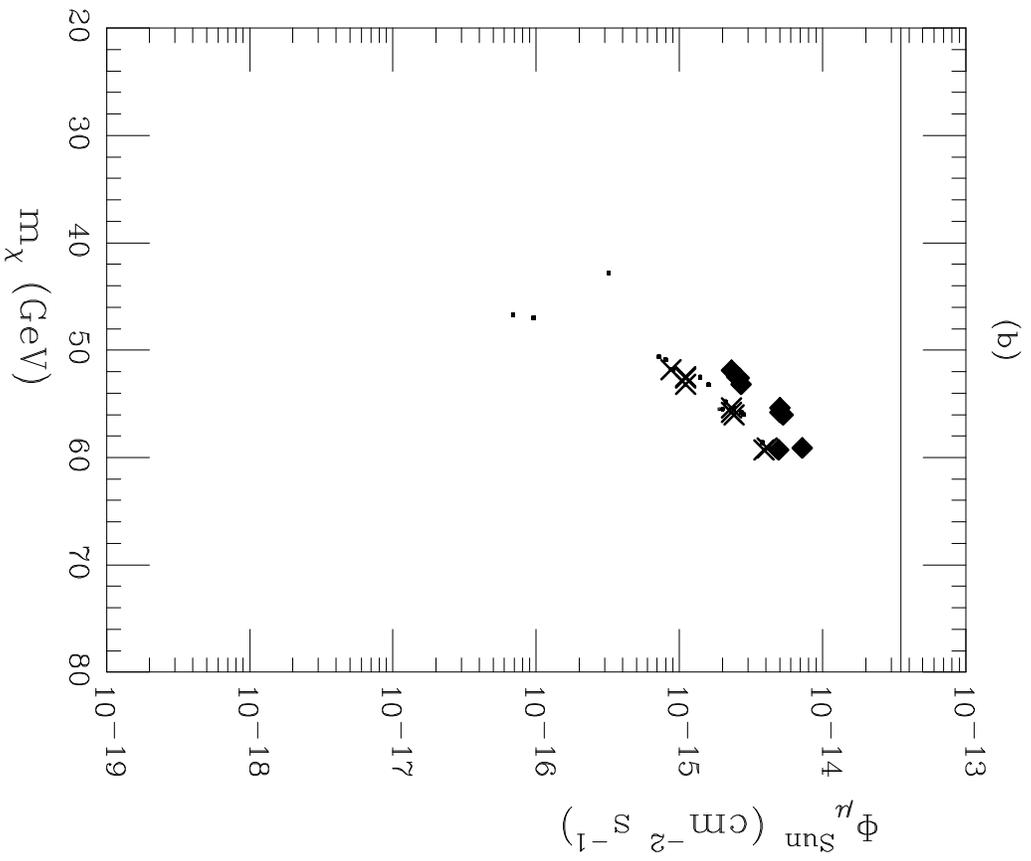
{\bf Figure 6} --
The flux $\Phi_{\mu}^{Sun}$ for configurations  
in region A (B) is given in Sect.a (b).
Diamonds, crosses and dots are as in Fig.4. 
The horizontal line denotes the 
 Baksan upper limit: 
$\Phi_{\mu}^{Sun} \leq 3.5 \times 10^{-14} {\rm cm^{-2} s^{-1}} (90 \% C.L.)$ 
\cite{bbb}.

\vspace{10 mm}

{\bf Figure 7} --
The flux $\Phi_{\mu}^{Earth}$ for configurations in region B. 
Diamonds, crosses and dots are as in Fig.4. 
The horizontal line denotes the 
 Baksan upper limit: 
$\Phi_{\mu}^{Earth} \leq 2.1 \times 10^{-14} {\rm cm^{-2} s^{-1}} (90 \% C.L.)$ 
\cite{bbb}.

\vspace{10 mm}

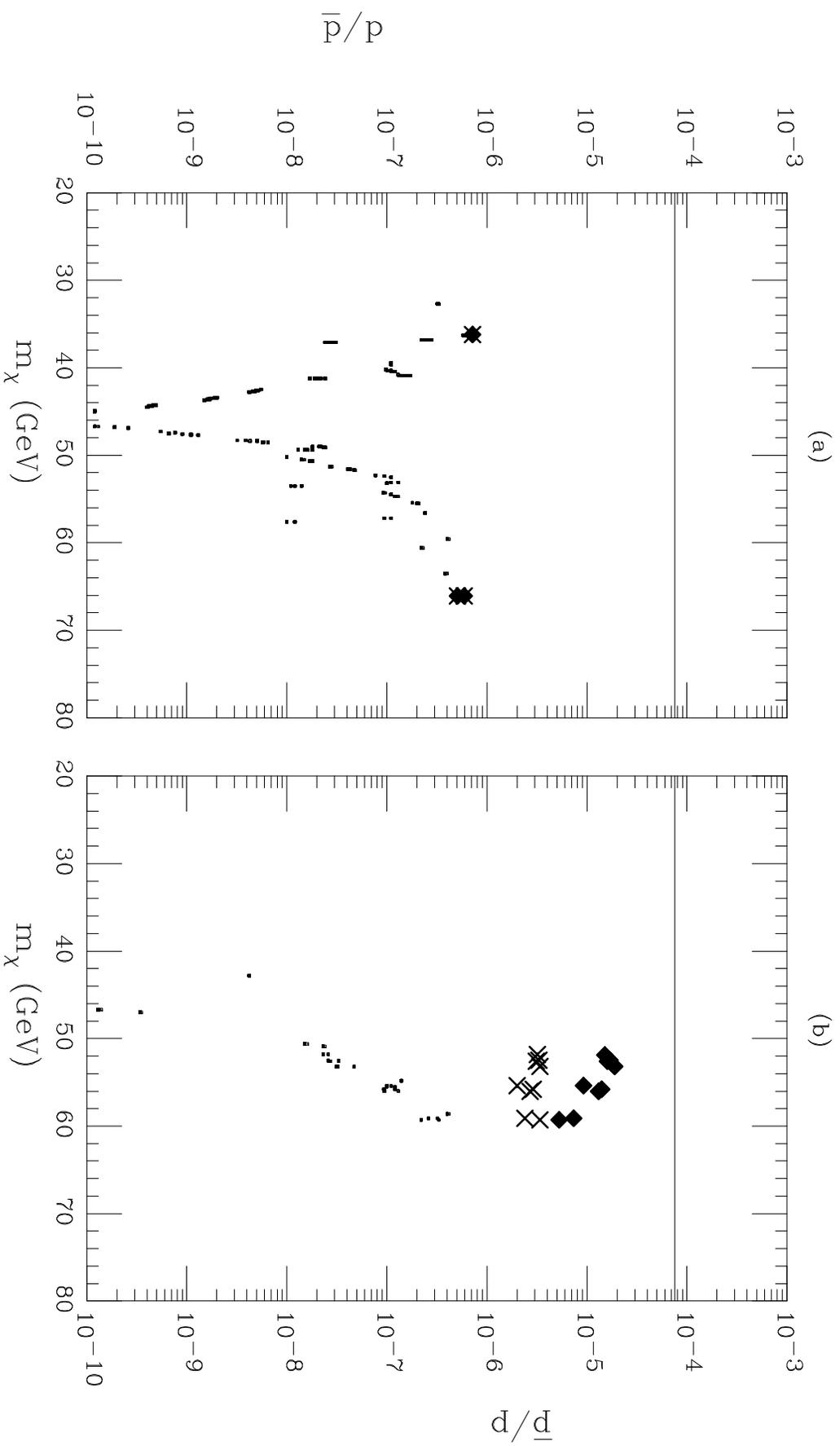
{\bf Figure 8} --
$\bar p/p$ ratio in cosmic rays due to neutralino-neutralino 
annihilation in the galactic halo. The energy integration range is 
given in the text. The scatter plot refers to configurations of region A 
(B) in Sect. a (b). The horizontal line corresponds to the 90 \% C.L. 
bound: 
$\bar p/p \leq 7.5 \times 10^{-5}$. 
Diamonds, crosses and dots are as in Fig.4.

\vspace{10 mm}

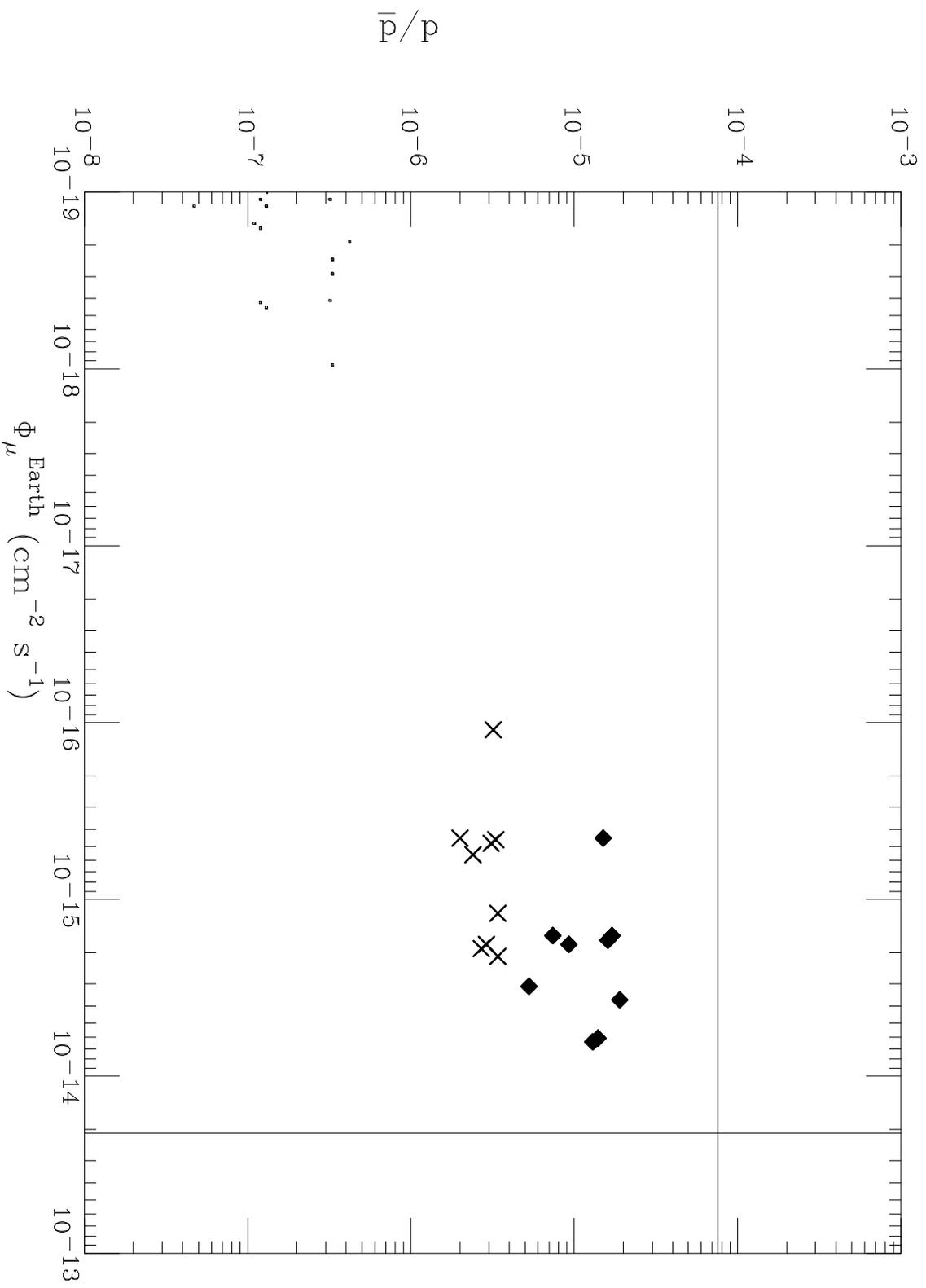
{\bf Figure 9} --
$\bar p/p$ versus $\Phi_{\mu}^{Earth}$ for configurations of region B. 
Diamonds, crosses and dots are as in Fig.4. 
The horizontal line corresponds to the 90 \% C.L. 
bound: 
$\bar p/p \leq 7.5 \times 10^{-5}$, the vertical line 
denotes the 
 Baksan upper limit: 
$\Phi_{\mu}^{Earth} \leq 2.1 \times 10^{-14} {\rm cm^{-2} s^{-1}} (90 \% C.L.)$.

\vspace{10 mm}

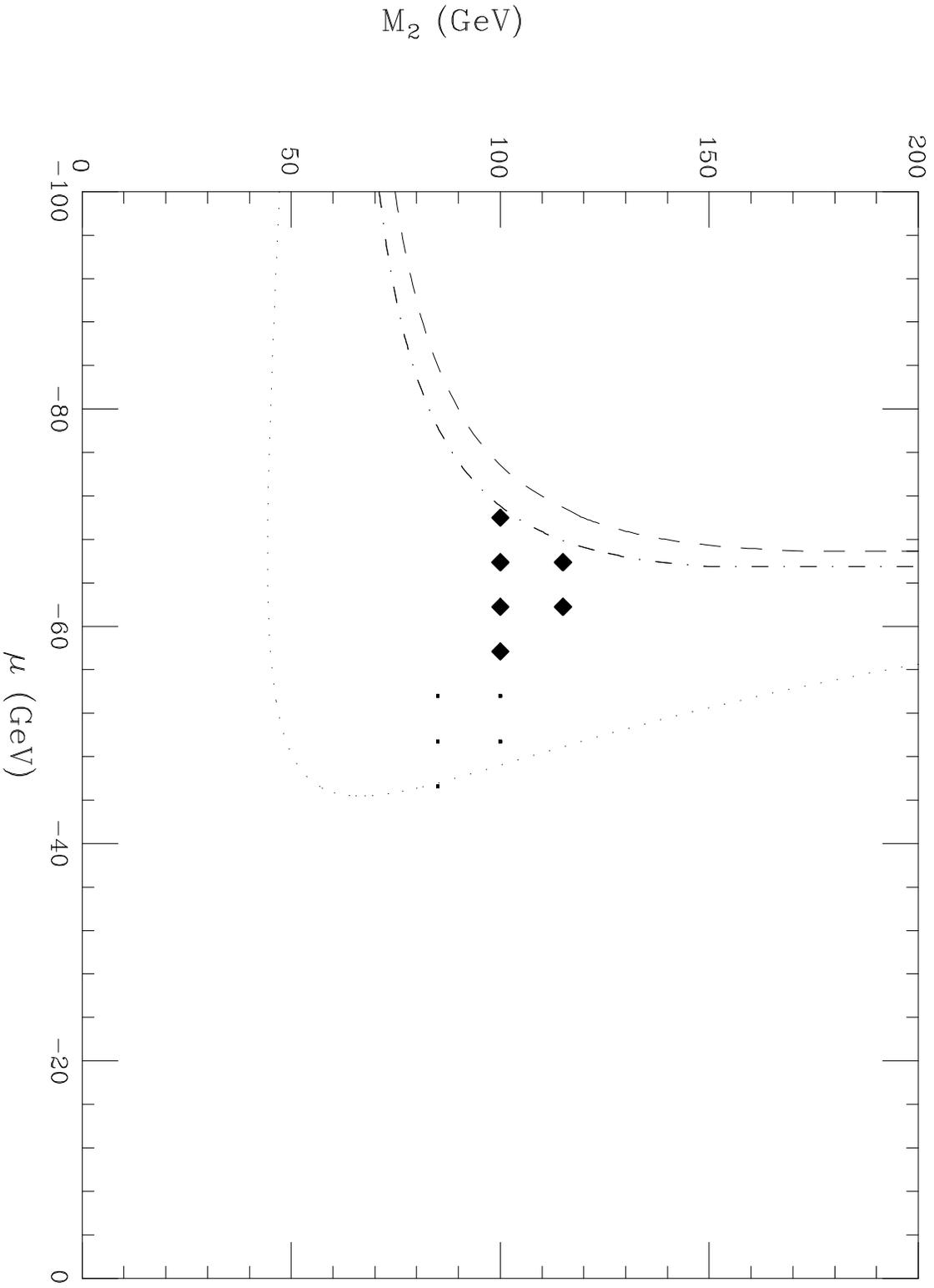
{\bf Figure 10} --
Location of the configurations 
of set S in the plane $\mu-M_2$ . The curves are denoted as in Fig.1. 
Diamonds denote the configurations of set S, dots the other allowed
configurations.


\begin{thebibliography}{99}
\bibitem{foot} These considerations apply to direct detection and to 
detection of neutrinos from the center of the Earth. 
The low-flux signals of neutrinos from the Sun which are  
related to spin-dependent effects will only be accessible to large 
area neutrino telescopes (AMANDA, NESTOR) whose threshold energy will 
however be too high to allow detection of $\chi_{AKM}$.
\end{thebibliography}
\end{document}